\documentclass[letterpaper]{article}
\usepackage{aaai2026}
\usepackage{times}
\usepackage{helvet}
\usepackage{float}
\usepackage{courier}
\usepackage[hyphens]{url}
\usepackage{graphicx}
\usepackage{natbib}
\usepackage{caption}
\usepackage{booktabs}
\usepackage[nolist]{acronym}
\usepackage{float}
\usepackage{amsmath}

\definecolor{copper}{HTML}{C84E00}
\definecolor{dandelion}{HTML}{FFD960}
\definecolor{piedmont}{HTML}{A1B70D}
\definecolor{eno}{HTML}{339898}
\definecolor{shale}{HTML}{0577B1}
\definecolor{dukeblue}{HTML}{012169}
\definecolor{ironweed}{HTML}{993399}
\definecolor{whisper}{HTML}{F3F2F1}
\definecolor{gingerbeer}{HTML}{FCF7E5}

\usepackage{tcolorbox}

\newtcolorbox{callout}{
    colback=gingerbeer!80,
    colframe=black,
    coltext=black,
    boxrule=1pt,
    arc=5pt,
    left=3pt,
    right=3pt,
    top=2pt,
    bottom=2pt,
}

\usepackage{booktabs}
\usepackage[nolist]{acronym}

\title{Ordinary, Reasonable Chatbots: Do AI Models Track Human Legal Judgments?}

\author{
    Nirav Patel\textsuperscript{\rm 1},
    Emily Wenger\textsuperscript{\rm 1},
    Christopher Buccafusco\textsuperscript{\rm 2}
}
\affiliations{
    \textsuperscript{\rm 1}Duke University, Departments of Computer Science \& Electrical and Computer Engineering\\
    \textsuperscript{\rm 2}Duke University Law School\\
}

\begin{document}

\maketitle

\begin{abstract}
As people increasingly rely on artificial intelligence (AI) for guidance in their own lives, scholars, lawyers, and even judges have begun to consider the role of AI in legal decision-making. As ``silicon sampling"\textemdash the use of generative AI models in social science research\textemdash is now impacting academia, ``silicon jurors" could make an appearance in courtrooms. This study joins an emerging line of research on generative AI models' ability to simulate human legal judgments. In particular, we study how large language model (LLM)-powered chatbots respond to series of questions about legal reasonableness. When the law needs to judge the appropriateness of a behavior, it most often asks whether the behavior was “reasonable.” Yet despite the ubiquity of reasonableness judgments, they are the site of constant vexation for lawyers, judges, and lay people. Reasonableness seems inherently vague and unpredictable, since it relies on variable context and implicit conceptual schemas. Moreover, many scholars caution that reasonableness judgments may vary along demographic lines. We compare the answers of human participants to those of twenty-six LLMs across twenty-five different legally relevant reasonableness judgments. Overall, our findings suggest that chatbot responses generally track those of human participants. Nonetheless, we find some suggestive\textemdash and potentially concerning\textemdash results. Compared to humans, LLMs generate more homogeneous responses and occasionally treat a variable standard as an invariant rule. And, compared to humans, LLMs tend to generate answers that are more favorable to the government and to corporations. Finally, our results indicate that LLMs’ responses tend to align more closely with those of respondents who are white, male, older, and more educated. More systematic research is needed to confirm or reject these initial findings.
\end{abstract}

\begin{acronym}

\end{acronym}

\section{Introduction}
\label{sec:introduction}
The popularity and fluency of generative AI chatbots like ChatGPT and Claude has led people to increasingly use these tools to answer a variety of different questions\textemdash about health, careers, relationships, and beyond. Given the perceived helpfulness and convenience of using chatbots in these domains,  scholars, journalists, and even federal judges have contemplated the possibility that AI could aid in various legal decision-making tasks and even, potentially, replace juries. Just as “silicon sampling” is being pushed in the social sciences~\cite{argyle2023out, dillion2023can, bisbee2024synthetic}, “silicon jurors” may soon begin to impact the law~\cite{grimm2024ai}.

In this paper, we contribute to the emerging research on AI’s ability to simulate human legal judgments~\cite{posner2025judge}. To do so, we assess how AI chatbots respond to a series of questions about legal “reasonableness.” When the law seeks to regulate the behavior of individuals or parties, the standard it most often reaches for is “reasonableness.” Variations from a contract's terms will be judged by their reasonableness. Government actions will often be subject to a standard of reasonableness across many fields. And most paradigmatically, torts defendants are generally required to exercise ordinary reasonable care.

Reasonableness judgments represent an exciting opportunity to study AI behavior for at least two important reasons. First, reasonableness is a legal “standard” rather than a “rule.” That is, determining what is reasonable is a matter of judgment, not simply the identification of some explicit attribute. Standards are vague in ways that rules are not. Second, for decades, scholars have argued that people’s judgments of reasonableness may vary along demographic lines. The seemingly objective reasonableness standard, they suggest, may preference a particular version of behavior as normatively desirable while ignoring other accounts~\cite{chamallas2018will,kahan2006cultural}. 

Studying how chatbots respond to prompts that call for undefined answers that may vary demographically will help us evaluate their use in a variety of important contexts, from answering laypeoples’ questions to supplementing legal or judicial decision-making. Our results have implications both for emerging legal issues as well as the wider scholarship on silicon sampling.
\paragraph{Our Contributions:}  
 \begin{itemize}
     \item  We compare how twenty-six LLM models from Meta, Google, Anthropic, OpenAI, DeepSeek, and xAI respond to judgments about the reasonableness of legally relevant scenarios. We compare the models’ responses to those of $500$ human participants who were asked about the same scenarios. 
     \item We find that LLM response distributions are statistically distinguishable from human response distributions, yet often remain concentrated within the empirical range of human responses, broadly tracking human notions of reasonableness.
     \item Further analysis reveals that models tend to generate answers that are more favorable to the government and corporations and align more closely with those of respondents who are white, male, older, and more educated.
 \end{itemize}


\section{Why and How Reasonableness Matters to Law}
\label{sec:human_reas}
Virtually every area of the law incorporates a standard of reasonableness or its opposite, unreasonableness. There is perhaps no more widely used term across legal doctrines than “reasonable.” Here, we briefly review the law’s uses of reasonableness and the debates about its meaning. We also explore recent research measuring lay people’s reasonableness judgments and whether reasonableness judgments may vary along demographic lines. 

\subsection{Reasonableness in the Law}
What does it mean to be reasonable, legally speaking? To explore the law’s use of the concept it is perhaps easiest to think about its paradigmatic use in tort (personal injury) law. Generally speaking, a defendant will have to compensate a victim for the harm they caused when the defendant’s conduct was negligent, which means that the defendant failed to exercise reasonable care when undertaking some activity, such as maintaining the safety of its premises. The defendant isn’t required to exercise extraordinary care, nor is its responsibility only minimal care.

Consider, for example, the reasonable amount of time for a grocery store to clean up a spill. The first thing to notice about this inquiry is that it is often vague. The law doesn’t demand that grocery stores clean up spills within three minutes or else face liability. It requires them to exercise reasonable care under the circumstances. In legal theory, this is the difference between using a “rule” and using a “standard.” Rules specify the conditions of compliance: the speed limit is 45 miles per hour; poisons must be labeled; the President of the United States must be at least thirty-five years old~\cite{kaplow2013rules}. By contrast, standards require the exercise of judgment based on other criteria. What is reasonable for one party to do might be unreasonable for another party, or circumstances might mean that what was once reasonable no longer is. As Benjamin Zipursky explains, “A legal provision operates as a standard rather than a rule when the capacity to apply the provision turns on the ability to utilize a relatively broader set of capacities for judgment, typically including purposive reasoning”~\cite{zipursky2014reasonableness}.

But how should these judgments be made? On one hand, we might determine whether the store was reasonable by considering some statistical or empirical account of grocery store behavior. We might want to know how long it takes most stores or the average store to clean up a spill. Separately, we might answer it by contemplating how long it \textit{should} take, whether or not that is how long it typically takes. That is, determining reasonable care is not simply a matter of counting but rather a normative judgment about appropriate or inappropriate behavior. 

Importantly, while many uses of the term ``reasonable" in the law are meant to reflect, at least in part, some empirical measure of lay human behavior, attitudes, or beliefs, other uses of the term may not. For example, antitrust law's prohibition on ``unreasonable" restraints of trade does not necessarily reflect any empirical or descriptive behavioral component nor is it meant to closely map ordinary citizens' views of anticompetitive behavior~\cite{areeda2020antitrust}. While some of the questions below represent the latter, most of the questions in our study involve issues where lay judgment is relevant. 

\subsection{Empirical Reasonableness and Demographic Variation}
Recently, scholars have begun surveying laypeople about their reasonableness judgments~\cite{jaeger2020empirical, kahan2006cultural, tobia2018people}. Following the experimental literature on normality~\cite{bear2017normality}, Tobia hypothesized that laypeople’s reasonableness judgments would fall between their beliefs about the average amount and the ideal amount of some quantity~\cite{tobia2018people}. Thus, people’s judgment of the reasonable number of weeks that a building project would be delayed would fall between their judgments of the average number of delayed weeks and the ideal number of delayed weeks. He surveyed over two hundred participants on thirteen different legally relevant scenarios and confirmed his hypothesis~\cite{tobia2018people}. Below, we use Tobia’s original thirteen questions plus twelve new questions as the foundation for our study. 

Whichever theoretical account of reasonableness one accepts, virtually all agree that laypeople’s reasonableness judgments matter. But empirically studying people’s reasonableness judgments is also important in light of concerns that, while the reasonableness standard projects neutrality, it is inevitably biased or variable, at least in certain contexts~\cite{alicke2021reasonable}. For example, scholars have suggested that men’s and women’s judgments of the reasonableness of a mistake about consensual sexual touching may differ~\cite{chamallas2018will,shoenfelt2002reasonable}. In addition, Black people’s views of the reasonableness of a police stop might differ from white people’s~\cite{bloom2023objective}.

\section{Why Study AI Reasonableness Judgements?}
\label{sec:ai_reas}
Given the potential complexity and variability of human reasonableness judgments, and in light of increasing interest in AI decision-making, understanding whether AI models track human judgments about an uncertain legal standard is valuable. How well do AI responses match those given by humans, and, to the extent that they differ, do they do so in systematic ways? These questions matter both for researchers interested in legal decision-making and AI’s role in it and for those studying the developments of and variance between AI models. 

\subsection{AI and Legal Judgment}
Reasonableness judgments present a fascinating opportunity to study AI decision-making for a number of reasons. First, there is increasing interest in AI’s capacities to track and predict human judgment. Some scholars anticipate a time in the not-too-distant future when AI decision-making will be able to replace, or at least supplement, human decision-making in legal settings~\cite{cardoso2024generative, winter2022challenges}. If LLMs’ legal judgments generally mimic humans’, the time and expense of litigation could be substantially diminished. In a recent opinion, Judge Kevin Newsom of the 11th Circuit Court of Appeals wrote: ``Might LLMs be useful in the interpretation of legal texts?  Having initially thought the idea positively ludicrous, I think I’m now a pretty firm `maybe.'"~\cite{snell_case}.

Studying AI reasonableness judgments is also interesting because reasonableness judgments involve the application of standards rather than rules. Answers to these questions cannot simply be arrived at from a search of relevant documents. Rather, humans rely on various schemas, norms, and empirical criteria when determining whether conduct is reasonable. Figuring out how these judgments are made\textemdash when only focused on human decision-makers\textemdash is already an enormously complex challenge, because both the inputs and their relative weights are largely unknowable. Thus, to the extent that AI judgments correspond to human judgments, the AIs have achieved an impressive conceptual task. Recent research by Arbel provides initial evidence that chatbots do, in fact, reflect some human conceptual schemas when making legal decisions~\cite{arbel2025silicon}.

\subsection{AI Homogeneity}
Recent evidence suggests that LLMs exhibit systematic homogeneity in open-ended generation, both within a single model’s repeated outputs and across different models’ outputs~\cite{bommasani2022picking,jiang2025artificial}. Jiang et al. formalize this concern with a large-scale evaluation of open-ended prompts, reporting an “Artificial Hivemind” effect characterized by within-model repetition and even stronger across-model homogeneity in responses. That is, models converge to strikingly similar responses on open-ended prompts~\cite{jiang2025artificial}. Similarly, Wenger and Kenett show that in standardized creativity tasks, LLM responses are substantially more similar to one another than human responses are to each other, even after controlling for variations in response structure~\cite{wenger2025we}.

Collectively, this evolving branch of research suggests homogeneity may be a broader property of LLMs. This is likely amplified by the possibility that many frontier LLMs draw from overlapping web-scraped “data commons,” especially large, shared corpora such as those derived from Common Crawl  \cite{vu2025happens}. More broadly, because LLM training objectives reward the efficient capture of statistical regularities and benchmark-optimized behavior, LLMs may naturally compress toward a relatively small set of high-probability response modes \cite{cropley2025cat,huh2024position}. This tendency is further reinforced by the widespread use of shared architectural paradigms \cite{kim2025correlated} (e.g., transformer-based models), similar pretraining objectives (next-token prediction), and overlapping fine-tuning and alignment procedures across model families \cite{li2025exploring,vu2025happens}. As a result, even independently trained models may converge on similar internal representations and output patterns when faced with open-ended prompts. 

In contrast, human judgment is intrinsically heterogeneous, shaped by diverse life histories, values, cultural backgrounds, and contextual priors that introduce variability that is not optimized away by a common objective function. This asymmetry suggests that homogeneity in LLM responses is not merely incidental, but a structural consequence of contemporary model design and training practices.

\section{Study Design}
\label{sec:metho}  
We study whether LLMs’ answers to legal reasonableness questions match human answers and whether they diverge in systematic ways. Because these questions do not have uniquely correct numerical answers, our focus is on the comparative distribution of human and model judgments rather than on legal correctness or any specific reasoning process. We therefore elicit models’ responses without additional factual context, retrieval, or sophisticated prompting scaffolds. The study consists of two principal parts. First, a survey of human participants to elicit their judgments of the reasonable or ideal amount of twenty-five legally relevant scenarios, and second, prompting LLMs with just the reasonable portion of the human survey, asking for their reasonable amount of the same twenty-five scenarios.

\subsection{Study Questions}
We pose questions about twenty-five legally relevant scenarios that are or might be governed by a reasonableness standard in the law, which we deployed to a sample population of $500$ humans (see Appendix Table~\ref{tab:demographics} for demographic coverage details) and twenty-six LLMs (see Appendix Table~\ref{tab:models_list} for model coverage details). The questions cover scenarios that are significant to a range of different legal fields, including torts, contracts, immigration, employment, criminal procedure, and family law. We adopted all thirteen of the questions that Tobia used in his survey \cite{tobia2018people}, with an additional twelve questions to expand the categories tested. 

Although scholars have increasingly studied AI models' abilities to engage in sophisticated legal reasoning tasks using complex and contextual scenarios~\cite{steging2025parameterized,guha2023legalbench,zheng2021does,chalkidis2022lexglue}, we have intentionally chosen a series of simple, low-context prompts requesting numerical estimates of legally reasonable quantities. While this choice has drawbacks, it also has benefits. First, it aligns with the existing empirical literature on legal reasonableness. Second, these questions are similar to the ones that laypeople are likely to ask language models, at least in the first instance. Third, and most importantly, our approach provides clear, easily interpretable assessments of models’ central tendencies and response distributions. We ask participants to provide the reasonable or ideal:

\begin{enumerate}
\setlength{\leftmargin}{0cm} 
\item  number of days taken to accept a business contract when no deadline is specified
\item  number of weeks taken to return a product ordered online when the warranty does not specify
\item  number of hours taken to reflect on an exciting but risky business proposition
\item  amount of unexpected additional costs in a \$10{,}000 building contract
\item  number of weeks that a building construction project is delayed beyond its stated completion date
\item  number of loud events held at a football field close to a quiet neighborhood, per year
\item  percent of profits that a car manufacturer spends on additional safety features
\item  percent of available medical details that a patient wants to hear from his/her doctor
\item  number of weeks that a person has to wait before being tried for a criminal charge
\item  number of dollars per hour that a charity pays in attorney's fees for legal work for the charity
\item  number of hours of notice that a landlord provides a tenant before entering the unit for maintenance or repairs
\item  interest rate for a loan
\item  percent likelihood that a company found legally liable for pollution will pollute again in the future
\item  number of inappropriate comments made by a coworker before they should be reported to a supervisor
\item  number of minutes that a typical police traffic stop should take
\item  amount of monthly balance someone should carry on a credit card with a \$30{,}000 limit
\item  number of days the government can detain an immigrant to remove them from the country
\item  number of weeks the police can use a location tracker like GPS on a crime suspect’s care without a warrant
\item  price that a city can charge for a permit to host a demonstration in a public park
\item  additional percentage of a restaurant bill that can be added for a ``living wage fee'' for kitchen staff
\item  difference between the mileage of a used car whose odometer reads 150{,}000 miles and the actual mileage
\item  number of yearly data breaches in a technology company
\item  number of hours 12 year old child can be left home alone
\item  number of dating partners that a divorced parent can introduce their 10 year old child to in a year
\item  number of months an employer can restrict a former employee from working for its competitors with a non-compete agreement
\end{enumerate} 

    A randomized subset of human participants were asked to give the {\em ideal} quantity, while the rest were asked to give a {\em reasonable} one. This was because, if LLMs varied from humans, we wanted to know whether they tended toward or away from the ideal. Both the human reasonable and LLM respondents were given the following initial text to guide their responses to the above questions:
\textit{Below, we ask you to estimate the reasonable quantity of a number of different things. Please note that you are not in any way being evaluated on these judgments, and we ask that you do not consult outside sources. In the following question we ask you to judge the legally reasonable quantity of a number of different things. We ask you to imagine that you are making these judgments in a legal setting for a legal purpose.} Human ideal respondents had a similar prompt but with ``reasonable'' replaced with ``ideal.''

\subsection{Human Participants Survey}
\label{sec:humans}

After receiving institutional IRB approval for our study design\footnote{Study approved by the Duke Univeristy IRB Protocol 2026-0109}, we recruited 500 participants using the online platform Prolific to complete a survey composed of the above prompt and questions. Participants were paid an average of \$12.57 per hour, prorated to completion time, and completed the survey and a brief demographic questionnaire in an average of 6 minutes and 41 seconds. Approximately half of our participants were asked about the reasonable amount of these behaviors, while the other half were asked for the ideal amount.

Consistent with prior work on legal reasonableness judgments, we applied outlier trimming at the question level rather than excluding entire respondents. This approach preserves statistical power by retaining otherwise valid participants while preventing a small number of extreme values from dominating question-specific estimates. Because several questions exhibited sparse, long-tailed response distributions that inflated the mean and standard deviation, we used the standard 1.5xIQR (inter-quartile range) rule to identify and remove extreme outliers. Importantly, this procedure is nonparametric and scale-free, making it well-suited to the heterogeneous units used across our questions (e.g., days, dollars, percentages), as it does not assume normality. This trimming step is conservative with respect to our subsequent hypothesis tests. Human responses are naturally more dispersed than model outputs, so removing extreme human outliers reduces heteroskedasticity and improves the comparability of human and model distributions without artificially increasing alignment. We use nonparametric statistical tests in our methodology to circumvent the impact of outlier handling procedures (retained observations per question after trimming are reported in Appendix Table~\ref{tab:question_metadata_n}). 

\subsection{Surveying AI Models}
\label{sec:llms}
 Using model APIs, we directly prompted twenty-six LLMs through a custom stateless pipeline that iterated over models and prompts. Models were provided the same context as humans via the system prompt with [Don't provide any reasoning or thinking, just provide your reasonable quantity without units.] appended. We therefore sampled each model–question pair twenty times to estimate each model's distribution of reasonable values. Prompting parameters were held constant across models. In particular, decoding settings such as temperature and top-p were fixed at their default values of 1 to ensure that observed differences across models reflect model behavior rather than parameter variation, while still permitting stochastic sampling from each model’s full token distribution.

For some model outputs, minor post-processing was applied to extract numeric unitless responses and to remove malformed outputs. When numeric responses could not easily be extracted, those observations were recorded as Not-A-Number (NaN). All post-processing steps were applied consistently across models. Models, like humans, produced some extreme outliers. For consistency, we applied the same 1.5×IQR trimming procedure to model responses as was used for the human data. In questions where the IQR was zero, we applied a 3$\sigma$ fallback rule to retain as much information as possible. As discussed in the methodology and results section, this preprocessing step does not materially affect the validity or power of our statistical tests (effective sample size for each model and question is reported in Appendix Table~\ref{tab:question_metadata_n}).

\subsection{Research Questions}

Each of the twenty-five reasonableness questions yields a distribution of human and LLM responses, collected via the metholodogy of \S\ref{sec:humans} and \S\ref{sec:llms}. From these responses, we study two key research questions. These questions, and our corresponding analysis tools, are described below. 

\begin{callout}
    \textbf{RQ1:} Are models “human-like” in their responses to judgements on reasonableness at the question level?
\end{callout} 

\paragraph{RQ1 Analysis Tools.}

\begin{figure*}[ht!]
\centering
\includegraphics[width=\textwidth]{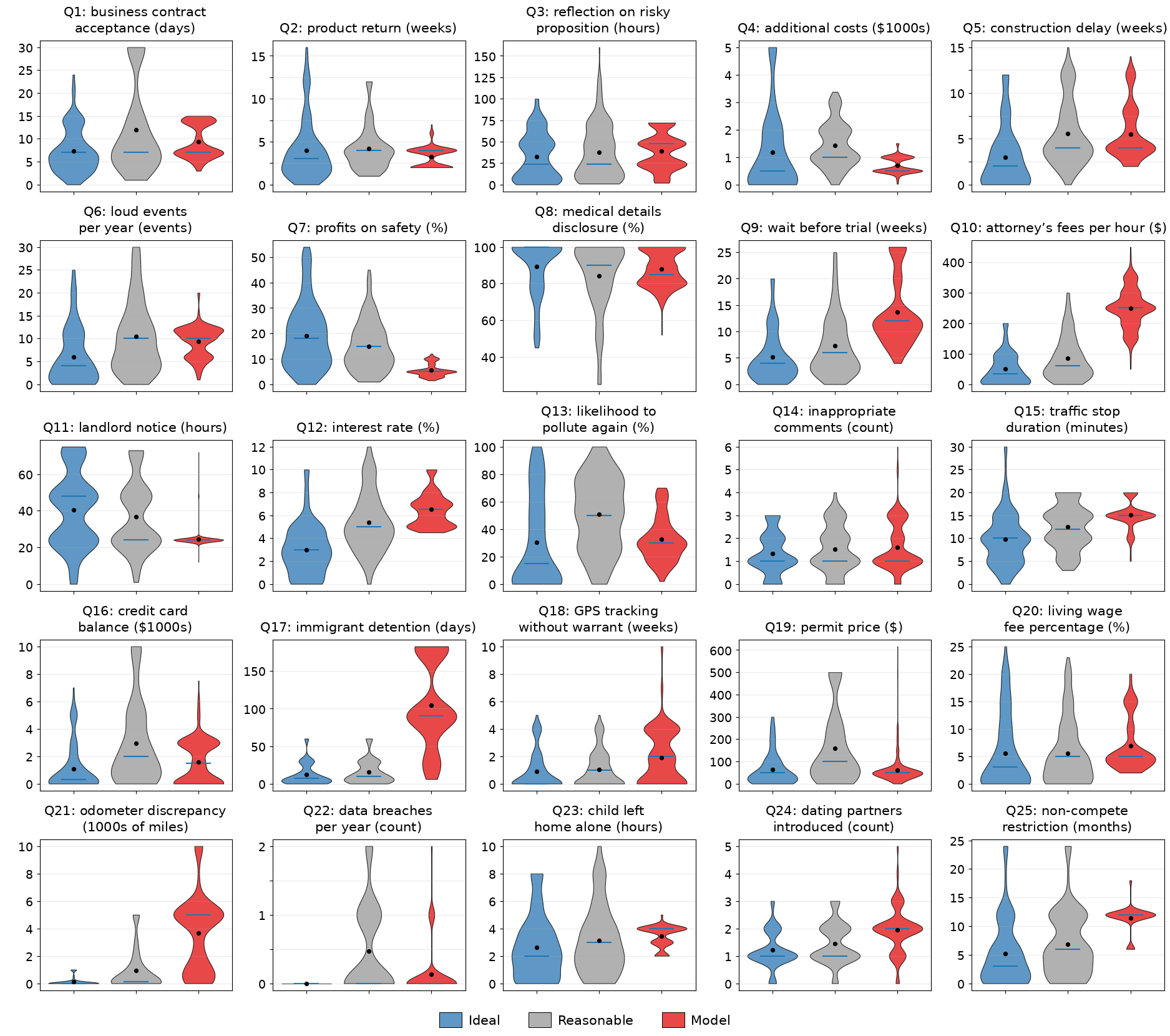}
\vspace{-0.26cm}
\caption{\textbf{Response Distributions by Question}, median is DASH line and mean is DOT.}
\label{fig:violins}
\vspace{-0.3cm}
\end{figure*}

To answer RQ1, we compare the distributions of model and human responses using three complementary statistical tests designed to capture distinct aspects of distributional similarity: dispersion (Brown–Forsythe), overall distributional divergence (Kolmogorov–Smirnov), and directional stochastic dominance (Brunner–Munzel). These tests are well-suited to our data, which exhibit skew and multimodality (see Figure~\ref{fig:violins}), violating the assumptions required for mean-based parametric methods. All three tests are valid under unequal variances and sample sizes. The tests are described in detail below. 

{\em Brown--Forsythe test for heteroscedasticity.}
To test whether models and humans differ in response variability, we apply the Brown--Forsythe test. For each group $g \in \{\text{human}, \text{model}\}$ with observations $\{y_{gi}\}_{i=1}^{n_g}$, we compute the group median $
\tilde{y}_g = \operatorname{median}(y_{g1}, \dots, y_{gn_g}),$ 
and then form absolute deviations $
z_{gi} = \lvert y_{gi} - \tilde{y}_g \rvert$.
The Brown--Forsythe statistic is obtained by performing a one-way ANOVA on $\{z_{gi}\}$ across the two groups, testing whether the mean absolute deviation differs between humans and models. The null hypothesis is human and model variances are equal, and the alternative hypothesis is that they differ, e.g. 
$H_0: \sigma_{\text{human}}^2 = \sigma_{\text{model}}^2,
H_1: \sigma_{\text{human}}^2 \neq \sigma_{\text{model}}^2.$

{\em Kolmogorov--Smirnov test for distributional equality.}
To test whether model and human responses are drawn from the same underlying distribution, we use the two-sample Kolmogorov--Smirnov test, which compares the empirical cumulative distribution functions (ECDFs) of the two groups. Let $F_H(x)$ and $F_M(x)$ denote the ECDFs of the human and model samples, respectively. The test statistic is $D = \sup_x \lvert F_H(x) - F_M(x) \rvert,$
which measures the maximum vertical distance between the two ECDFs. Large values of $D$ indicate systematic differences in the location, spread, or shape of the two distributions. The null hypothesis is that the two samples come from the same empirical distribution, and the alternative hypothesis is that they come from different empirical distributions, e.g.
$H_0: F_H(x) = F_M(x) \forall x,
H_1: \exists\, x \text{ such that } F_H(x) \neq F_M(x).$

{\em Brunner--Munzel test for stochastic dominance.}
To assess whether models (or humans) tend to give systematically larger responses than humans (or models), we use the Brunner--Munzel test. This test evaluates the stochastic dominance parameter $
p = P(M > H) + \tfrac{1}{2} P(M = H),$
where $M$ and $H$ denote random draws from the model and human response distributions, respectively. This quantity represents the probability that a randomly drawn model response exceeds a randomly drawn human response, with ties split evenly. Values of $p > 0.5$ indicate that models tend to give larger responses than humans, while $p < 0.5$ indicates the reverse ($H_0: p = 0.5, H_1: p \neq 0.5$). 

For each test, we apply the procedure independently to each question (twenty-five tests total) and control family-wise error rate across questions with a Bonferroni correction.

\begin{callout}
    \textbf{RQ2:} Do models exhibit any systematic variation from humans in their responses to judgments on reasonableness across questions?
\end{callout}

\paragraph{RQ2 Analysis Tools.}

 To answer RQ2, we first identify demographic dimensions with meaningful variation among humans, then assess whether model predictions align more closely with one subgroup than another using regression. Additionally, we examine cross-question variation patterns using the cosine similarity metric.

{\em Demographic analysis.} We evaluate demographic variation across several stages. First, we coarsen each demographic variable into two primary groups in order to create more balanced comparisons while preserving substantively meaningful distinctions. These binary groupings allow us to conduct statistical comparisons with adequate sample sizes while still reflecting interpretable demographic contrasts. Gender categories with very low coverage are excluded at this stage because their sample sizes are too small to support reliable statistical comparison.

Age is collapsed into 18–44 years old, and 45+ years old. Ethnicity is recoded into White (self-reported “White” only) and Non-White (all other responses, including multiracial and non-White categories). Education is collapsed into Associates or less and Bachelors or more. Gender is analyzed using Man and Woman as the two comparison groups. These binary demographic groups are used consistently throughout the analysis, both in the per-question regression models (see question-level regression analysis in Appendix) and in the cross-question cosine-similarity embeddings. In parallel, model responses are grouped by model family, allowing us to evaluate demographic alignment against families of models.

{\em Cosine similarity for across question reasonable judgment bias.} 
Finally, we examine cross-question alignment between models and humans using cosine similarity. We represent each group—human demographic subgroups and model families—as a vector in a 25-dimensional space, where each dimension corresponds to one reasonableness question. Because the questions differ in units and scale, we apply per-question min–max scaling jointly across all human and model vectors prior to computing similarities. This normalization ensures that each question contributes comparably and that cosine similarity reflects relative response patterns across questions rather than raw numeric magnitude. In this way, alignment captures whether models and humans exhibit similar profiles of judgment across the full set of scenarios, rather than agreement driven by a small number of high-variance or large-scale questions.
For humans, we construct group-level response profiles by taking both the mean and the median across respondents within each demographic subgroup (age, gender, education, and ethnicity). For models, we compute analogous vectors at the model-family level by aggregating responses across all models belonging to each family. These group-level vectors summarize typical response patterns while abstracting away individual-level noise.

We then compute cosine similarity between each model-family vector and each human subgroup vector, using both mean-based and median-based representations. This yields a measure of how closely each model family’s pattern of reasonableness judgments aligns with those of specific human demographic groups across the entire 25-question survey.

We substantiate whether this cosine similarity matrix reflects meaningful demographic structure by performing a Monte Carlo sampling procedure over the scaled response space. For each demographic split (gender, ethnicity, age, and education), we construct empirical distributions of cosine similarities by repeatedly sampling survey response vectors for both models and each human demographic group, drawing each response independently from each question pool (see Appendix Figure~\ref{fig:montecarlo}). This produces paired normal distributions of model–human cosine similarities for each subgroup, allowing us to estimate the expected alignment between models and each demographic group. We then compare these distributions using two-sample t-tests on their means, assessing whether models exhibit systematically different levels of alignment across demographic groups.

\begin{table*}[t]
\centering
\small
\resizebox{0.99\textwidth}{!}{
\begin{tabular}{lccccccl}
\toprule
& \multicolumn{2}{c}{\textbf{Brown--Forsythe}} & \multicolumn{2}{c}{\textbf{Kolmogorov--Smirnov}} & \multicolumn{3}{c}{\textbf{Brunner--Munzel}} \\
\cmidrule(lr){2-3} \cmidrule(lr){4-5} \cmidrule(lr){6-8}
Question & $W$ & Significance & $D$ & Significance & $P(M>H)$ & Direction & Significance \\
\midrule
Q1: business contract acceptance (days) & 67.01 & *** & 0.22 & *** & 0.51 & Model $>$ &  \\
Q2: product return (weeks) & 40.18 & *** & 0.19 & *** & 0.41 & Human $>$ & ** \\
Q3: reflection on risky proposition (hours) & 30.17 & *** & 0.23 & *** & 0.55 & Model $>$ &  \\
Q4: additional costs (\$) & 148.49 & *** & 0.44 & *** & 0.22 & Human $>$ & *** \\
Q5: construction delay (weeks) & 5.21 &  & 0.14 &  & 0.52 & Model $>$ &  \\
Q6: loud events per year (events) & 186.08 & *** & 0.25 & *** & 0.52 & Model $>$ &  \\
Q7: profits on safety (\%) & 453.42 & *** & 0.59 & *** & 0.17 & Human $>$ & *** \\
Q8: medical details disclosure (\%) & 61.73 & *** & 0.17 & ** & 0.50 & Human $>$ &  \\
Q9: wait before trial (weeks) & 0.15 &  & 0.49 & *** & 0.79 & Model $>$ & *** \\
Q10: attorney’s fees per hour (\$) & 7.20 &  & 0.75 & *** & 0.94 & Model $>$ & *** \\
Q11: landlord notice (hours) & 376.11 & *** & 0.44 & *** & 0.32 & Human $>$ & *** \\
Q12: interest rate (\%) & 59.87 & *** & 0.40 & *** & 0.68 & Model $>$ & *** \\
Q13: likelihood to pollute again (\%) & 114.24 & *** & 0.41 & *** & 0.29 & Human $>$ & *** \\
Q14: inappropriate comments (count) & 0.92 &  & 0.08 &  & 0.52 & Model $>$ &  \\
Q15: traffic stop duration (minutes) & 140.07 & *** & 0.35 & *** & 0.66 & Model $>$ & *** \\
Q16: credit card balance (\$) & 57.08 & *** & 0.30 & *** & 0.36 & Human $>$ & *** \\
Q17: immigrant detention (days) & 117.40 & *** & 0.78 & *** & 0.94 & Model $>$ & *** \\
Q18: GPS tracking without warrant (weeks) & 48.73 & *** & 0.21 & *** & 0.62 & Model $>$ & *** \\
Q19: permit price (\$) & 186.72 & *** & 0.46 & *** & 0.32 & Human $>$ & *** \\
Q20: living wage fee percentage (\%) & 17.43 & *** & 0.38 & *** & 0.62 & Model $>$ & *** \\
Q21: odometer discrepancy (miles) & 66.46 & *** & 0.49 & *** & 0.82 & Model $>$ & *** \\
Q22: data breaches per year (count) & 79.52 & *** & 0.27 & *** & 0.36 & Human $>$ & *** \\
Q23: child left home alone (hours) & 230.63 & *** & 0.33 & *** & 0.61 & Model $>$ & ** \\
Q24: dating partners introduced (count) & 2.85 &  & 0.34 & *** & 0.67 & Model $>$ & *** \\
Q25: non-compete restriction (months) & 209.83 & *** & 0.54 & *** & 0.78 & Model $>$ & *** \\
\bottomrule
\end{tabular}
}
\caption{\textbf{Brown--Forsythe ($W$), Kolmogorov--Smirnov ($D$), and Brunner--Munzel ($P(M>H)$) statistics table} $^{***}$, $^{**}$, and $^{*}$ denote Bonferroni-corrected $p<0.001$, $p<0.01$, and $p<0.05$, respectively.}
\label{tab:stat_tests}
\end{table*}

\section{Key Findings}

\subsection{Result of RQ1 analysis}
\label{sec:rq1_results}

\begin{callout}
 In general, LLMs do a fairly good job of tracking human reasonableness judgments. Their responses tend to align with humans’ responses to a set of challenging legal questions. Nonetheless, LLMs exhibit some interesting differences: they are more prone to homogeneity and they appear to be more pro-government and pro-corporation.
\end{callout}

\paragraph{LLM/Human Alignment.} 
Table~\ref{tab:stat_tests} summarizes our three statistical comparisons between human and model responses for each of the twenty-five reasonableness questions. As described in the methodology, each test captures a distinct notion of statistical similarity between models and humans: the Brown–Forsythe test evaluates differences in response dispersion, the Kolmogorov–Smirnov test assesses global distributional equality, and the Brunner–Munzel test evaluates directional stochastic dominance. To best illustrate how to interpret the significant results, consider first the non-significant results of Q5 (Construction Delay) and Q14 (Inappropriate Comments) across all three tests. For these two questions, there is insufficient evidence to conclude that humans and models differ in spread, overall distribution, or response tendency. In Figure~\ref{fig:violins}, we visualize the distribution of responses across the Human Ideal, Human Reasonable, and Model Reasonable surveys using violin plots. Looking at these, the non-significant results of Q5 and Q14 are easy to confirm. The reasonable and model distributions overlap closely in both shape and central tendency.

To illustrate the Brown–Forsythe results, consider Q7 (Profits on Safety). The very large statistic (W=453.42) indicates an extreme and statistically significant difference in response dispersion, with human judgments exhibiting far greater variability than the tightly clustered model outputs. This reflects a substantive difference in within-group disagreement and is also clearly visible in the corresponding violin plot. To illustrate the Kolmogorov–Smirnov results, Q10 (Attorneys’ Fees) provides a particularly striking example. The KS statistic (D=0.75) indicates a large and statistically significant separation between the human and model distributions. That is, there exists a threshold at which the cumulative share of human and model responses differs by 75 percentage points, reflecting major differences in both distributional shape and location. Finally, to illustrate the Brunner–Munzel results, Q17 (Immigrant Detention), P(model$>$human)=0.94 indicates that model responses exceed human responses in approximately 94$\%$ of pairwise comparisons, revealing strong stochastic dominance of model judgments over human judgments for this question. This too is visible on the corresponding violin plot.

Though many questions exhibit statistically significant differences between human and model responses, the magnitudes of these differences vary substantially across the three tests. Out of twenty-five questions, twenty show significant Brown–Forsythe results, twenty-three show significant Kolmogorov–Smirnov results, and nineteen show significant Brunner–Munzel results after correction, with sixteen questions significant under all three tests. The Brown–Forsythe results frequently correspond to very large effects, reflecting the pronounced homogeneity of model outputs relative to the much broader spread of human judgments. This pattern is consistent with the tightly clustered model distributions observed in the violin plots and in the MAD table.

\begin{table}[t]
\centering
\small
 \resizebox{0.47\textwidth}{!}{
\begin{tabular}{lrr}
\toprule
Question & Model MAD & Human MAD \\
\midrule
Q9: wait before trial & \textbf{4.0} & \textbf{4.0} \\
Q16: credit card balance & \textbf{1500.0} & \textbf{1500.0} \\
Q22: data breaches per year & \textbf{0.0} & \textbf{0.0} \\
Q24: dating partners introduced & \textbf{0.0} & \textbf{0.0} \\
Q12: interest rate & \textbf{1.5} & 2.0 \\
Q18: GPS tracking without warrant & 2.0 & \textbf{1.0} \\
Q14: inappropriate comments & \textbf{0.0} & 1.0 \\
Q1: business contract acceptance & \textbf{2.0} & 3.0 \\
Q2: product return & \textbf{1.0} & 2.0 \\
Q5: construction delay & \textbf{0.5} & 2.0 \\
Q23: child left home alone & \textbf{0.0} & 2.0 \\
Q20: living wage fee percentage & \textbf{2.0} & 5.0 \\
Q6: loud events per year & \textbf{2.0} & 5.0 \\
Q7: profits on safety & \textbf{2.0} & 5.0 \\
Q15: traffic stop duration & \textbf{0.0} & 3.0 \\
Q3: reflection on risky proposition & 24.0 & \textbf{21.0} \\
Q10: attorney’s fees per hour & 50.0 & \textbf{40.0} \\
Q13: likelihood to pollute again & \textbf{10.0} & 25.0 \\
Q17: immigrant detention & 60.0 & \textbf{8.0} \\
Q19: permit price & \textbf{0.0} & 75.0 \\
Q4: additional costs & \textbf{0.0} & 500.0 \\
Q21: odometer discrepancy & 2000.0 & \textbf{100.0} \\
\bottomrule
\end{tabular}
}
\caption{\textbf{Median Absolute Deviations (MAD)}, more homogenous in bold.}
\label{tab:mad_sorted}
\vspace{-0.2cm}
\end{table}

By contrast, many of the statistically significant Brunner–Munzel and Kolmogorov–Smirnov results correspond to more moderate effect sizes. Among the nineteen questions with significant Brunner–Munzel results, only four exhibit extreme stochastic dominance (with P(model$>$human)$\geq$0.80 or $\leq$0.20), while the remaining significant cases reflect smaller but systematic directional shifts. Similarly, although twenty-three questions are significant under Kolmogorov–Smirnov, only four show very large separations (D$\geq$0.50), indicating that many distributional differences, while reliable, do not involve near-complete separation of the human and model distributions. This pattern underscores that models often differ from humans in response variability even when their central tendencies remain within the range of human judgments.

Despite applying outlier handling across all three distributions, long tails persisted in several of the model response distributions. Notably, this pattern reflects a lack of intrinsic variability in the models rather than genuine dispersion. For example, in Q19 (Permit Price), nearly all model responses fell between 0 and 100, with both the 25th and 75th percentiles equal to fifty, yielding an IQR range of 0. The same collapse of the IQR occurred for Q11 (Landlord Notice), Q15 (Traffic Stop Duration), Q22 (Data Breaches), Q23 (Child Alone), and Q24 (Dating Partners); in Q11 in particular, responses were almost perfectly homogeneous, with most models returning 24.

At the same time, a small number of extreme model outputs appear in the distribution tails. These sparse outliers are neither removed by the IQR rule, which becomes ineffective when the IQR collapses to 0, nor our fallback $3\sigma$ rule, which itself is sensitive to extreme tails. The result is violin plots with concentrated central mass and elongated, low-density tails, showing largely consistent model behavior with occasional outlier responses.

Crucially, the statistical comparisons we report are robust to these tail behaviors and to the inclusion or exclusion of extreme model outputs. Our primary tests, the Brunner–Munzel test, the Kolmogorov–Smirnov test, and the Brown–Forsythe test, are all nonparametric or rank-based and do not rely on assumptions of normality, equal variances, or well behaved means and standard deviations. Rather than being driven by a small number of extreme values, these tests are sensitive to differences in distributional shape, stochastic dominance, and spread across the full sample. As a result, the presence of sparse tail outliers in otherwise highly concentrated model distributions does not materially affect inference. The detected differences between human and model responses reflect systematic shifts in location and dispersion across the bulk of the distributions, not artifacts of a few extreme points.
\vspace{-0.1cm}
\paragraph{LLM Homogeneity.} We further explore this notion of LLM homogeneity using a Median Absolute Deviation (MAD) table (see Table~\ref{tab:mad_sorted}). MAD provides a robust, distribution-free measure of dispersion that quantifies how tightly responses cluster around a central value, making it well suited for our survey data. For a set of responses $\{y_i\}_{i=1}^n$, MAD is defined as $\operatorname{MAD} = \operatorname{median}_i \lvert y_i - \tilde{y} \rvert$, where $\tilde{y}$ is the sample median. We compute MAD separately for human and model responses on the reasonable questions. In this context, smaller MAD values indicate stronger agreement, meaning responses concentrate around a shared canonical judgment, while larger MAD values indicate greater disagreement and heterogeneity in what is considered reasonable. 
\vspace{-0.1cm}
\paragraph{Pro-Government and Pro-Corporation Alignment.} Finally, we observe several interesting tendencies in  the model results (see Table 1). First, the models’ responses tend to be meaningfully more supportive of the government than human responses. Thus, the models allow significantly longer detentions for traffic stops (Human median 12 vs LLM median 15 minutes), more time before a criminal defendant is brought to trial (median 6 vs 12 weeks), longer warrantless GPS tracking (median 1 vs 2 days), and longer detention of immigrants before deportation (median 10 vs 90 days). Second, the models may lean more towards corporate interests than do our human participants. For example, they support significantly less money spent on safety (Human median 15 vs LLM median 5 percent), they think corporations are less likely to pollute again (median 50 vs 30 percent), and they allow much longer non-compete restrictions (median 6 vs 12 months). Other questions are more ambiguous on this dimension (e.g. product returns and construction delays), warranting further study. 

\begin{figure*}[t]
\centering
\includegraphics[width=0.49\textwidth]{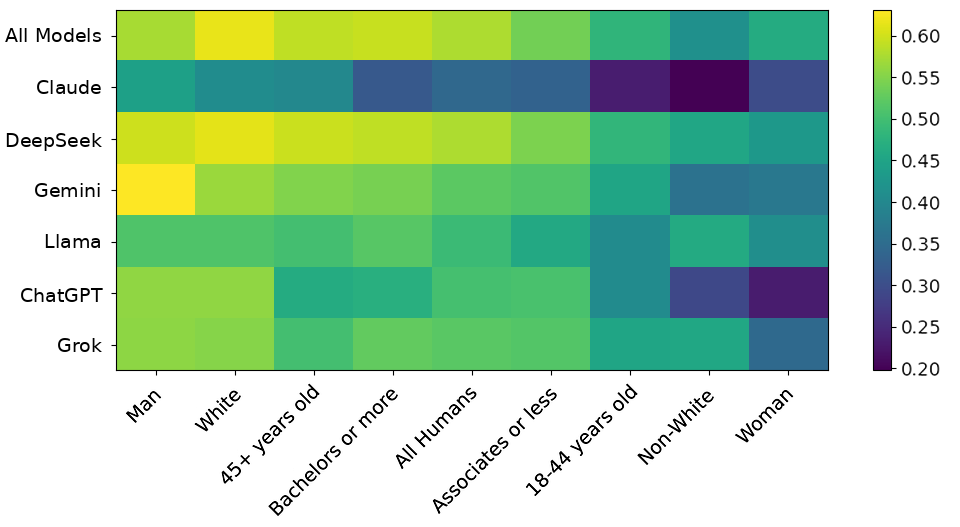}
\hfill
\includegraphics[width=0.49\textwidth]{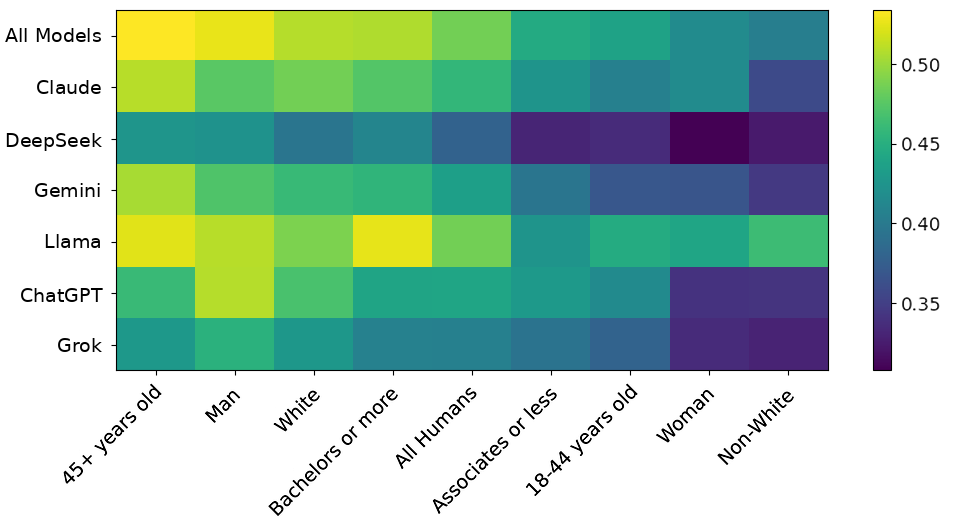}
\caption{\textbf{Model Family and Demographic Group Cosine Similarity}, using median (left) and mean (right) response vectors. Lighter cells indicate greater similarity.}
\label{fig:cosine_sim}
\end{figure*}

\begin{table*}[t]
\centering
\small
\begin{tabular}{llllccccc}
\toprule
Split & Group A & Group B & Mean$_A$ & SD$_A$ & Mean$_B$ & SD$_B$ & $t$ & Cohen's $d$ \\
\midrule
Gender & Man & Woman & 0.7020 & 0.0729 & 0.6945 & 0.0750 & 7.17$^{***}$ & 0.10 \\
Ethnicity & White & Non-White & 0.7020 & 0.0727 & 0.6856 & 0.0754 & 15.63$^{***}$ & 0.22 \\
Age & 45+ & 18--44 & 0.7056 & 0.0725 & 0.6909 & 0.0746 & 14.15$^{***}$ & 0.20 \\
Education & Bachelors+ & Associates or less & 0.7056 & 0.0737 & 0.6874 & 0.0735 & 17.52$^{***}$ & 0.25 \\
\bottomrule
\end{tabular}
\caption{\textbf{Independent two-sample $t$-tests}, comparing mean alignment scores across demographic groups.}
\label{tab:ttest_demographics}
\vspace{-0.2cm}
\end{table*}

\subsection{Result of RQ2 analysis}
\label{sec:rq2_results}
\begin{callout}
 Across the full range of reasonableness questions, LLMs respond more like older, white, male, and more educated human respondents than they do the complementary groups. 
\end{callout}

While regression results (see Appendix) indicate that strong, question-specific demographic alignment is relatively rare, they do not rule out the possibility that more subtle patterns of model leaning emerge when responses are considered in aggregate. Even when individual questions show only small or statistically insignificant differences, consistent shifts in the same direction across many questions could accumulate into meaningful demographic alignment. To examine whether such high-dimensional structure exists, we therefore turn to patterns of similarity across all twenty-five questions simultaneously.

Across questions, the cosine similarity analyses reveal consistent structure. In both the median-based and mean-based embeddings in Figure~\ref{fig:cosine_sim}, the response profiles of men, white respondents, more educated respondents, and older respondents are more similar to every model family than those of their complementary groups. The fact that this pattern appears in both plots suggests that it is not driven by outliers or distributional skew, but may suggest a systematic subtle alignment in the way models and these demographic groups respond across the full set of reasonableness judgments.

Since all model families exhibit broadly similar alignment gradients across demographic groups, we aggregate across models and compare their overall alignment to each human subgroup. Under this pooled view, the Monte Carlo cosine-similarity analysis reveals statistically detectable differences in mean alignment across all four demographic splits reported in Table~\ref{tab:ttest_demographics}. Models are, on average, more closely aligned with men than women, with white than non-white respondents, with older (45+) than younger (18–44) respondents, and with respondents holding a bachelor’s degree or higher than those with less education.

While the absolute shifts in cosine similarity are modest, the consistency of these differences across thousands of sampled model–human pairings indicates that model reasonableness judgments are not demographically neutral. Instead, they exhibit systematic variation in their alignment with different human groups. Moreover, because these comparisons are made in a 25-dimensional response space, even small per-question variations can compound across dimensions, leading to more substantial aggregate differences in how models encode and reproduce patterns of reasonableness across demographic groups. 

Future work should examine a broader range of models, alongside larger and more demographically diverse human samples. This would help determine whether the observed alignment patterns generalize beyond the models and datasets studied here and whether models preserve, amplify, or diminish demographic differences in human judgments.

\section{Discussion and Implications}
\label{sec:FW}

This study contributes to the emerging line of research studying generative AI decision-making in a variety of contexts, including especially the law \cite{albashrawi2025generative,winter2022challenges}. Legal reasonableness is an inherently variable and vague standard that typically lacks any connection to ground truth. Our principal finding is that AI models perform fairly well at open-ended legal decision-making tasks, at least to the extent of matching human participants’ judgments. Although we do note some suggestive divergences that are worthy of future research. 

Across a wide range of legal contexts, AI models generate responses that are similar to human responses. Although the models often provided answers that were statistically different from humans, our total impression is one of overall coherence. When looking across the range of questions in our survey, the models’ responses did a strong job of approximating the human responses–even for an inherently vague legal standard. For no question is the mean or median response from the models wildly divergent from the human response. Simple visual analysis of the violin plots in Figure~\ref{fig:violins} indicates that both the central tendencies of the models and their overall distribution of responses tend to match the human reasonableness responses. 

We stress that the answers to legal reasonableness judgment questions are unlikely to exist in LLMs' latent training data in the way that the answer to a legal question like “What is the minimum age for a U.S. Senator?” will be. Yet the models were able to generate answers that often approximated human responses. 
Despite the general success of the models at approximating human judgments, they did have some divergences that are worthy of sustained study. 

\paragraph{Model Homogeneity.} First, for some of the questions, the AI models demonstrate a tendency towards homogeneity that is not present in the human responses. Legally speaking, the models treat a legal standard like a legal rule. For example, on the question about the reasonable amount of notice a landlord must provide a tenant before entering their apartment, the models almost uniformly answer 24 hours. We see similar, if somewhat less stark, levels of homogeneity for the questions about profits spent on safety (Q7, 5$\%$), the price of a permit for a demonstration (Q19, $\$$50), data breaches per year (Q22, 0), and months of non-compete restrictions (Q25, 12). Again, all of these are governed by reasonableness standards rather than rules, yet the models respond more homogeneously than humans do. 

This finding has implications for emerging concerns about model homogeneity \cite{huh2024position,jiang2025artificial,wenger2025we}. While there might be many tasks for which we expect models to generate homogeneous answers, there are many others–including in the legal domain–where homogeneity may be costly. It might be problematic, for example, if model responses to some questions fail to represent the variation and heterogeneity of human judgments and instead proceed as if there were only one correct answer. This could be especially concerning in fields that require creativity and in settings where multiple responses could be suitable, such as legal or policy decisions. 

\paragraph{Model Tendencies and Demographic Comparisons.} Although the AI models tend to track human reasonableness judgments, our results may suggest some systematic differences, including pro-government and pro-corporation bias as noted in \S\ref{sec:rq1_results}, as well as our finding that LLMs align more closely with male, white, older, and more educated respondents than they do with complementary groups (\S\ref{sec:rq2_results}). 

While these findings are preliminary, they are suggestive of concerning patterns of LLM behavior. A substantial emerging literature studies the extent to which LLMs generate answers that lean towards one end of the political spectrum or the other. Our results run counter to some recent findings that AI chatbots tend to produce more “liberal” responses \cite{rozado2024political,schwenzow2023political}. One possible difference is that our questions are less obviously politically sensitive than those in other studies. But, because we do not know our respondents’ politics, it is also possible that our sample skews liberal. 

In future research, we plan to expand our demographic sample of human respondents, including by oversampling groups that are less well represented in our data. It would also be helpful to include specific questions about respondents' politics. To further study our findings about models’ pro-government and pro-corporation leanings, we plan to introduce more questions that address these issues. Finally, we plan to expand the LLM survey by adding a longitudinal component and by adding models, including the new legal research tool LLM Harvey. 
\section{Limitations}
\label{sec:limit}
Recent scholarship has begun to study AI legal decision-making in sophisticated, context-dependent scenarios~\cite{rotolo2025rule,bomfim2025automating,linna2026challenges}. Our aims are more modest. Because we are interested in understanding LLMs' central tendencies and distributions, we have necessarily sacrificed meaningful nuance required to evaluate legal reasoning. In addition, our questions were not conceived with the goal of detecting pro-government or pro-corporate responses from LLMs. Further research in this direction should employ queries that are more clearly targeted to study existing bias. And future work could use broader demographic sampling and a wider range of models to disentangle sources of variation in these preliminary findings.

\appendix
\newpage
\section*{Appendix}
\label{sec:appendix}


\begin{table}[h]
\centering
\small

\textbf{Age}

\begin{tabular}{lcc}
\toprule
Category & \% (Reasonable) & \% (Ideal) \\
\midrule
35--44 years old      & 28.3 & 31.2 \\
25--34 years old      & 23.9 & 23.3 \\
45--54 years old      & 22.3 & 21.7 \\
55--64 years old      & 15.0 & 10.3 \\
65+ years old         & 7.3  & 8.7  \\
18--24 years old      & 2.8  & 4.3  \\
Prefer not to answer & 0.4  & 0.4  \\
\bottomrule
\end{tabular}

\vspace{1em}
\textbf{Gender}

\begin{tabular}{lcc}
\toprule
Category & \% (Reasonable) & \% (Ideal) \\
\midrule
Man                     & 49.4 & 49.0 \\
Woman                   & 48.6 & 49.0 \\
Prefer not to answer    & 1.2  & 0.4  \\
Prefer to self-describe & 0.8  & 1.6  \\
\bottomrule
\end{tabular}

\vspace{1em}
\textbf{Ethnicity}
\resizebox{0.48\textwidth}{!}{
\begin{tabular}{lcc}
\toprule
Ethnicity & \% (Reasonable) & \% (Ideal) \\
\midrule
American Indian or Alaskan Native & 0.0 & 0.4 \\
Asian & 8.9 & 5.5 \\
Black or African American & 9.7 & 8.3 \\
Hispanic Latino, or Spanish Origin & 4.0 & 4.3 \\
White & 73.7 & 71.9 \\
Mixed & 3.2 & 8.7 \\
Prefer not to answer & 0.4 & 0.8 \\
\bottomrule
\end{tabular}
}

\vspace{1em}
\textbf{Education}

\resizebox{0.48\textwidth}{!}{
\begin{tabular}{lcc}
\toprule
Category & \% (Reasonable) & \% (Ideal) \\
\midrule
Grades 1 through 11 & 0.4 & 0.8 \\
12th grade—no diploma & 0.4 & 0.4 \\
GED or alternative credential & 0.4 & 0.4 \\
Regular high school diploma & 12.1 & 8.7 \\
Some college credit, less than 1 year & 8.1 & 9.5 \\
1 or more years of college credit, no degree & 6.5 & 9.9 \\
Associate's degree (AA, AS) & 12.6 & 12.6 \\
Bachelor’s degree (BA, BS) & 44.1 & 36.8 \\
Master’s degree (MA, MS, MBA, etc.) & 12.6 & 16.6 \\
Professional degree (MD, JD, etc.) & 1.6 & 1.2 \\
Doctorate degree (PhD, EdD) & 0.8 & 2.0 \\
Prefer not to answer & 0.4 & 1.2 \\
\bottomrule
\end{tabular}
}

\caption{\textbf{Demographic Distributions}}
\label{tab:demographics}
\end{table}


\begin{table*}[h]
\centering
\small
\begin{tabular}{llcc}
\toprule
Model name & Family & Year & Open source \\
\midrule
claude-3-5-haiku-latest & Claude & 2024 & False \\
claude-3-7-sonnet-latest & Claude & 2024 & False \\
claude-opus-4-0 & Claude & 2025 & False \\
claude-opus-4-1 & Claude & 2025 & False \\
claude-sonnet-4-0 & Claude & 2025 & False \\
claude-sonnet-4-5 & Claude & 2025 & False \\
deepseek\_deepseek-r1 & DeepSeek & 2025 & True \\
deepseek\_deepseek-r1-0528 & DeepSeek & 2025 & True \\
deepseek\_deepseek-v3-0324 & DeepSeek & 2025 & True \\
gemini-2.0-flash & Gemini & 2025 & False \\
gemini-2.0-flash-lite & Gemini & 2025 & False \\
gemini-2.5-flash & Gemini & 2025 & False \\
gemini-2.5-flash-lite & Gemini & 2025 & False \\
gemini-2.5-pro & Gemini & 2025 & False \\
meta\_llama-3.3-70b-instruct & Llama & 2024 & True \\
meta\_llama-4-maverick-17b-128e-instruct-fp8 & Llama & 2025 & True \\
meta\_llama-4-scout-17b-16e-instruct & Llama & 2025 & True \\
meta\_meta-llama-3.1-405b-instruct & Llama & 2024 & True \\
meta\_meta-llama-3.1-8b-instruct & Llama & 2024 & True \\
openai\_gpt-4.1 & OpenAI & 2025 & False \\
openai\_gpt-4.1-mini & OpenAI & 2025 & False \\
openai\_gpt-4.1-nano & OpenAI & 2025 & False \\
openai\_gpt-4o & OpenAI & 2024 & False \\
openai\_gpt-4o-mini & OpenAI & 2024 & False \\
xai\_grok-3 & xAI & 2025 & False \\
xai\_grok-3-mini & xAI & 2025 & False \\
\bottomrule
\end{tabular}
\caption{\textbf{Models Metadata}}
\label{tab:models_list}
\end{table*}


\begin{table*}[h]
\centering
\small
\begin{tabular}{lccc}
\toprule
Short Label (Units) & Reasonable N & Ideal N & Models \\
\midrule
Q1: business contract acceptance (days)        & 236 & 216 & 392 \\
Q2: product return (weeks)                     & 226 & 224 & 515 \\
Q3: reflection on risky proposition (hours)    & 240 & 237 & 518 \\
Q4: additional costs (\$)                      & 225 & 243 & 507 \\
Q5: construction delay (weeks)                 & 223 & 237 & 514 \\
Q6: loud events per year (events)              & 233 & 238 & 514 \\
Q7: profits on safety (\%)                     & 223 & 235 & 493 \\
Q8: medical details disclosure (\%)            & 224 & 225 & 509 \\
Q9: wait before trial (weeks)                  & 229 & 227 & 470 \\
Q10: attorney’s fees per hour (\$)             & 228 & 234 & 502 \\
Q11: landlord notice (hours)                   & 238 & 244 & 517 \\
Q12: interest rate (\%)                        & 231 & 246 & 496 \\
Q13: likelihood to pollute again (\%)          & 247 & 252 & 497 \\
Q14: inappropriate comments (count)            & 232 & 239 & 508 \\
Q15: traffic stop duration (minutes)           & 223 & 252 & 509 \\
Q16: credit card balance (\$)                  & 232 & 226 & 448 \\
Q17: immigrant detention (days)                & 219 & 239 & 515 \\
Q18: GPS tracking without warrant (weeks)      & 224 & 223 & 496 \\
Q19: permit price (\$)                         & 228 & 213 & 507 \\
Q20: living wage fee percentage (\%)           & 240 & 243 & 514 \\
Q21: odometer discrepancy (miles)              & 223 & 211 & 509 \\
Q22: data breaches per year (count)            & 222 & 206 & 505 \\
Q23: child left home alone (hours)             & 239 & 243 & 414 \\
Q24: dating partners introduced (count)        & 240 & 247 & 510 \\
Q25: non-compete restriction (months)          & 233 & 240 & 493 \\
\bottomrule
\end{tabular}
\caption{\textbf{Questions and retained sample sizes for reasonable, ideal, and model responses.}}
\label{tab:question_metadata_n}
\end{table*}

\subsection*{Question Level Demographic and Model Family Regression Analysis}

\textit{Regression for question-level reasonable judgment bias.}
For each question and each demographic split, we first conduct Brunner--Munzel tests for stochastic dominance, correcting for multiple comparisons across questions using a Bonferroni adjustment. We restrict subsequent regression analyses to the subset of question--demographic pairs that exhibit statistically significant differences, in order to avoid overfitting and unnecessary multiple testing.

For each retained case, we estimate a linear regression in which the AI prediction is treated as the baseline outcome and demographic subgroup indicators capture deviations from that baseline:
\[
Y_i = \beta_0 + \beta_1 D_{1i} + \beta_2 D_{2i} + \varepsilon_i,
\]
where $Y_i$ is the model’s predicted value for observation $i$, $\beta_0$ represents the model-family baseline prediction, and $D_{1i}$ and $D_{2i}$ are binary indicators for the two demographic subgroups under comparison.

To assess whether AI predictions align more closely with one subgroup than the other, we compare squared deviations from the model baseline by conducting one-sided Wald tests on $\beta_1^2$ and $\beta_2^2$. Because these squared coefficients are nonlinear functions of the estimated parameters, their standard errors are computed using the delta method. Formally, we test
\[
H_0: \beta_1^2 \ge \beta_2^2
\qquad \text{versus} \qquad
H_1: \beta_1^2 < \beta_2^2,
\]
which evaluates whether the model’s prediction is statistically closer to subgroup~1 than to subgroup~2.

These regressions are estimated separately for each model family, and the resulting $p$-values are corrected for multiple testing using a Bonferroni adjustment.

As an initial screening step, we test for directional differences between coarsened human demographic groups using the Brunner–Munzel test, which evaluates whether one subgroup tends to give systematically larger responses than another. Across the 25 questions, significant subgroup differences emerge in only three cases: Q10 (Attorneys’ Fees), where respondents with a Bachelor’s degree or higher give higher values than those with less education; Q19 (Permit Price), where men give higher values than women; and Q12 (Interest Rate), where men also give higher values. These results confirm that demographic differences in human responses are relatively limited in this dataset, in-line with existing literature.\cite{tobia2018people}

We then restrict our regression analysis to these three cases to avoid overfitting and unnecessary multiple testing. Of these, only Q10 and Q12 yield statistically meaningful model alignment effects. For Q10 (Attorneys’ Fees), all model families produce predictions that are significantly closer to the responses of the more educated group, indicating systematic alignment with higher education. For Q12 (Interest Rate), DeepSeek, Gemini, Llama, and Claude produce predictions that are significantly closer to men’s responses than women’s. In contrast, Q19 (Permit Price) does not exhibit statistically reliable model alignment with either gender, despite the presence of a human subgroup difference.


\begin{figure*}[h]
\centering
\includegraphics[width=\textwidth]{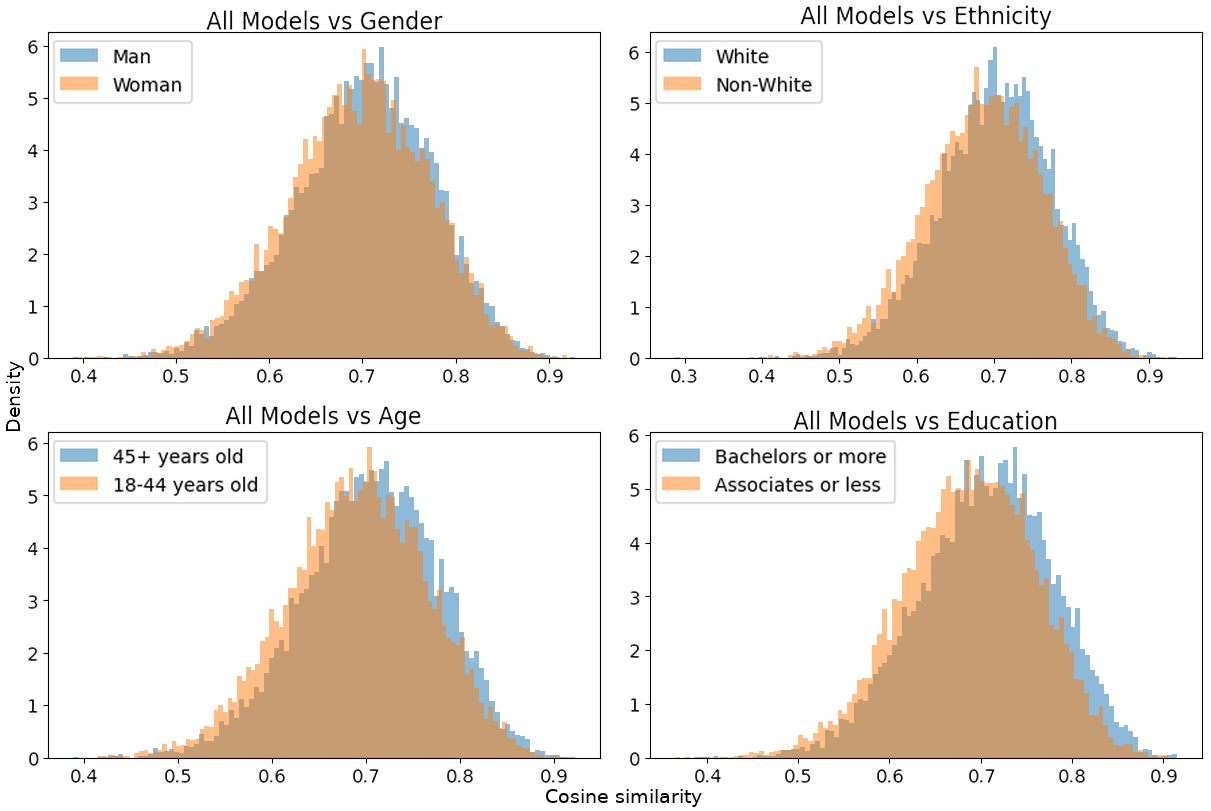}
\caption{\textbf{Repeated Cosine Similarity Monte Carlo Distributions}}
\label{fig:montecarlo}
\end{figure*}
\clearpage
\clearpage
\section*{Ethical Statement}
\label{sec:ethi}
IRB approval for the human participant study was obtained, and participants signed a clearly written consent form. Participant data was anonymized and stored on secure servers. Other ethical risks from this paper are minimal, as our LLM survey does not involve sensitive data and elicits benign model responses.

\section*{Acknowledgments}
We thank Soumya Ganti, Tamesha Derkach, and Jake McAuliffe for excellence research assistance, our colleagues in the Duke Law School for their
suggestions of relevant literature, and the anonymous reviewers for their helpful feedback.

\bibliography{MyCite}

\end{document}